\documentclass[fleqn,usenatbib]{mnras}

\usepackage{times}
\usepackage[T1]{fontenc}
\usepackage{graphicx}
\usepackage{amsmath}
\usepackage{amssymb}
\usepackage{flushend}

\newcommand{\Msunnom}{\hbox{$\mathcal{M}^{\rm N}_\odot$}}
\newcommand{\Rsunnom}{\hbox{$\mathcal{R}^{\rm N}_\odot$}}

\newcommand{\Teff}{\ensuremath{T_{\rm eff}}}                      
\newcommand{\kms}{\,km\,s$^{-1}$}                                 

\newcommand{\reff}[1]{#1}
\newcommand{\refff}[1]{#1}

\title[Four bright eclipsing binaries with $\gamma$\,Doradus stars]
      {Four bright eclipsing binaries with $\gamma$\,Doradus pulsating components: CM\,Lac, MZ\,Lac, RX\,Dra and V2077\,Cyg}

\author[Southworth \& Van Reeth]
       {John Southworth\,$^{1}$, Timothy~Van Reeth\,$^2$ \\
        $^1$\,Astrophysics Group, Keele University, Staffordshire, ST5 5BG, UK \\
        $^2$\,Institute of Astronomy, KU Leuven, Celestijnenlaan 200D, B-3001 Leuven, Belgium
        }

\date{Accepted YYYYMMDD. Received YYYYMMDD; in original form YYYYMMDD.}

\begin{document} \label{firstpage} \pagerange{\pageref{firstpage}--\pageref{lastpage}} \maketitle 

\begin{abstract}
The study of pulsating stars in eclipsing binaries holds the promise of combining two different ways of measuring the physical properties of a star to obtain improved constraints on stellar theory. Gravity (g) mode pulsations such as those found in $\gamma$\,Doradus stars can be used to probe rotational profiles, mixing and magnetic fields. Until recently few $\gamma$\,Doradus stars in eclipsing binaries were known. We have discovered g-mode pulsations in four detached eclipsing binary systems from light curves obtained by the Transiting Exoplanet Survey Satellite (TESS) and present an analysis of their eclipses and pulsational characteristics. We find unresolved g-mode pulsations at frequencies 1--1.5\,d$^{-1}$ in CM\,Lac, and measure the masses and radii of the component stars from the TESS data and published radial velocities. MZ\,Lac shows a much richer frequency spectrum, including pressure modes and tidally-excited g-modes. RX\,Dra is in the northern continuous viewing zone of TESS so has a light curve covering a full year, but shows relatively few pulsation frequencies. For V2077\,Cyg we formally measure four pulsation frequencies, but the available data are inadequate to properly resolve the g-mode pulsations. V2077\,Cyg also shows total eclipses, with which we obtain the first measurement of the surface gravity of the faint secondary star. All four systems are bright and good candidates for detailed study. Further TESS observations are scheduled for all four systems, with much improved temporal baselines in the cases of RX\,Dra and V2077\,Cyg.
\end{abstract}



\section{Introduction}
\label{sec:intro}

Eclipsing binary star systems (EBs) are crucial objects for understanding the physics governing stellar structure and evolution, because they are the only stars whose masses and radii can be measured to high precision and accuracy from observational material and geometrical arguments alone. Precise measurements of the masses and radii of stars in EBs were instrumental in the development of stellar theory \citep[e.g.][]{Russell14obs}, in the verification of the first modern generation of theoretical stellar models \citep[e.g.][]{Andersen++90apj,Pols+97mn}, and continue to be used to guide theoretical progress \citep{ClaretTorres18apj,Tkachenko+20aa}. In the current era of \'echelle spectroscopy and space-based light curves it is possible to measure masses and radii for stars in EB to precisions approaching 0.1\% \citep{Maxted+20mn,Graczyk+21aa}. The impact of photometry from space missions has been reviewed in detail by \citet{Me21univ}.

Another class of stars capable of posing high-quality constraints on stellar theory is that of the pulsating variables \citep{Aerts++10book}. Among these, stars showing gravity-mode (g-mode) pulsations are well-suited to studying the interiors of stars as g-modes, which have buoyancy as the dominant restoring force, can travel deep within stars whilst leaving observable signatures on the stellar surface \citep{Bowman20faas}. $\gamma$\,Doradus ($\gamma$\,Dor) variables \citep{Kaye+99pasp} are stars of spectral types A and F that show g-mode pulsations with periods ranging from 0.3\,d to 4\,d and amplitudes up to 0.1\,mag \citep{Henry++07aj,Grigahcene+10apj}.

In recent years, asteroseismic analyses of main-sequence stars with g-mode pulsations have allowed us to place constraints on multiple phenomena. These include near-core stellar rotation \citep[e.g.][]{Bouabid+13mn,Vanreeth+16aa,Christophe+18aa,Takata+20aa,Takata+20aa2,Szewczuk++21mn}, convective core boundary mixing \citep[e.g.][]{Michielsen+19aa,WuLi19apj,Mombarg+21aa}, extra envelope mixing \citep[e.g.][]{Mombarg+20apj,Pedersen+21natas} and magnetic fields \citep[e.g.][]{Prat+19aa,Vanbeeck+20aa,Lecoanet+22mn}.

The advantages of asteroseismology and binarity can be combined in objects which show both eclipses and pulsations. These systems have the potential to set exacting constraints on stellar theory \citep[e.g.][]{Johnston++19aa}. A wide variety of pulsating stars are known in EBs, including $\delta$\,Scuti \citep{Maceroni+14aa,Dasilva+14aa,Lee+20pasj,Lee+21aj,Me21obs6}, $\beta$\,Cephei \citep{Me+20mn,Me++21mn,LeeHong21aj}, SPB \citep{Clausen96aa,MeBowman22mn} and $\delta$\,Cephei \citep{Pilecki+18apj}. Over the last decade, multiple $\gamma$\,Dor variables in EBs have been discovered as well. An emblematic case is that of KIC\,11285625 \citep{Debosscher+13aa}. Others were found by, among others, \citet{Hambleton+13mn,Lee+14aj,Sowicka+17mn,Helminiak+17aa,Guo++17apj2,Lampens+18aa,Guo+19apj} and \citet{Vanreeth+22aa}.

A sample of 115 $\gamma$\,Dor variables in EBs was identified by \citet{GaulmeGuzik19aa} via a systematic search of the \textit{Kepler} EB catalogue \citep{Kirk+16aj}. \citet{Li+20mn} performed an asteroseismic evaluation of these targets and reported the detection of g-mode period-spacing patterns for 34 of them, as well as for one discovered by \citet{Colman+22apjs}. An independent systematic search and asteroseismic evaluation of the \textit{Kepler} EB catalogue was conducted by \citet{Sekaran+20aa}, resulting in a different sample of 93 $\gamma$\,Dor variables in EBs, with detected period-spacing patterns for seven of them. So far, asteroseismic modelling has only been achieved for a small number of targets: KIC\,10080943 \citep{SchmidAerts16aa,Johnston+19mn}, KIC\,10486425 \citep{Zhang+18apj2}, KIC\,7385478 \citep{GuoLi19ApJ} and KIC\,9850387 \citep{Zhang+20apj,Sekaran+21aa}.

The accumulating data from the NASA Transiting Exoplanet Survey Satellite (TESS; \citealt{Ricker+15jatis}) is enabling previously unattainable analyses of many of the bright and well-known variable stars. The catalogue of objects includes many EBs with a long observational history \citep[e.g.][]{Me20obs}. In the course of trawling this database we have discovered four bright EBs whose light curves also show $\gamma$\,Dor pulsations. In this work we present these discoveries, studies of their light and radial velocity (RV) curves, and a first analysis of the nature of their pulsations.


\section{Observations}
\label{sec:obs}

\begin{figure*}
\includegraphics[width=\textwidth,angle=0]{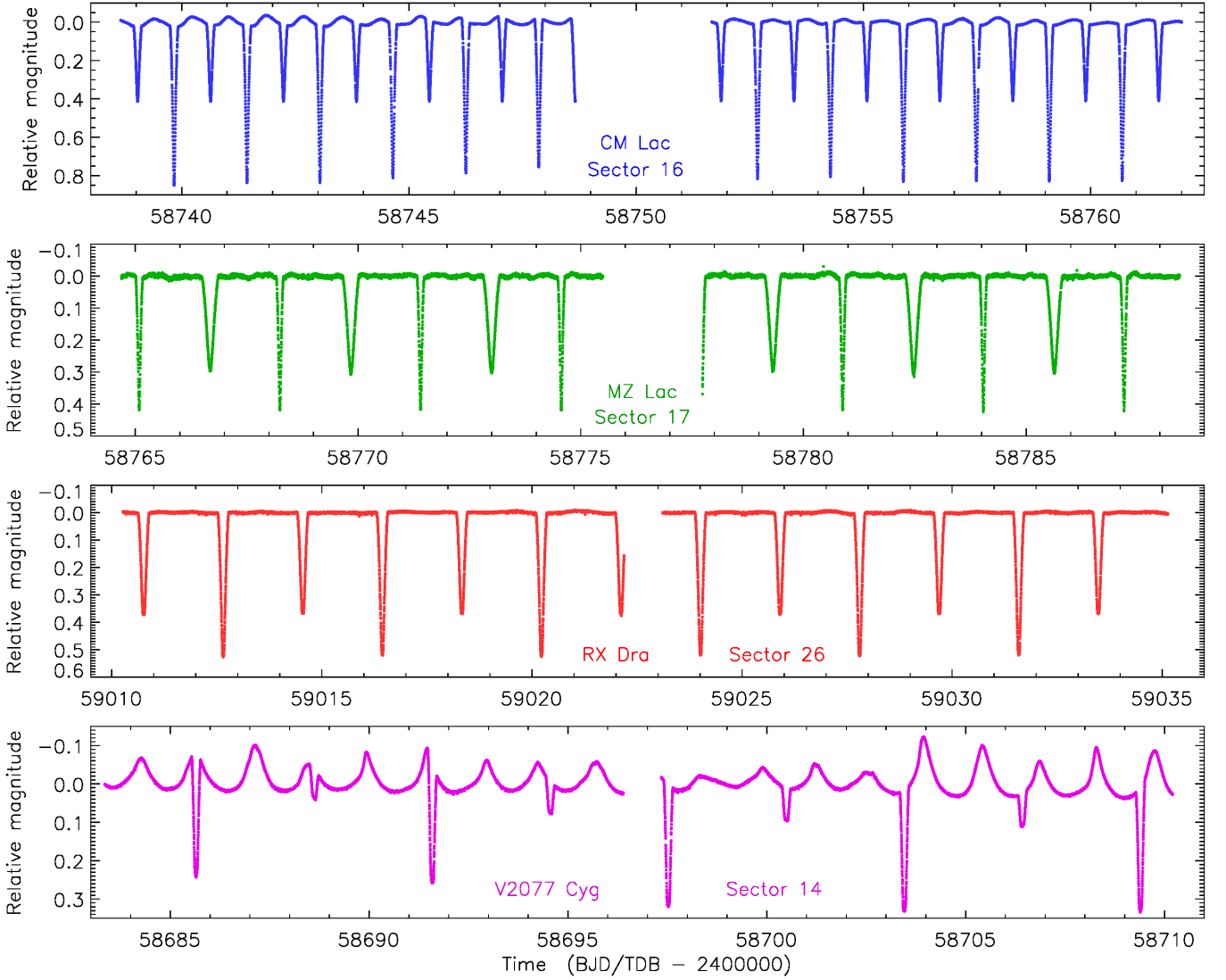}
\caption{\label{fig:time} TESS SAP light curves of the four EBs.
In each case only one sector is shown.} \end{figure*}

The TESS satellite \citep{Ricker+15jatis} is in the process of observing almost the entire celestial sphere, split into 69 overlapping sectors. Each sector is a 24$^\circ${$\times$}96$^\circ$ strip of sky and is photometrically monitored for 27.4\,d, through a filter with a high throughput between 600\,nm and 1000\,nm, with a pause near the midpoint for the download of data to Earth. During the nominal two-year mission, full-frame images (FFI) were taken at a default cadence of 1800\,s, while a subset of stars were monitored in short cadence, where successive exposures were combined to yield light curves with a cadence of 120\,s. In the ongoing extended mission, the FFI cadence has been changed to 600\,s. 

Reduced light curves are obtained from the data by the Science Processing Operations Center \citep[SPOC;][]{Jenkins+16spie} and made available through the MAST portal\footnote{\texttt{https://mast.stsci.edu/portal/Mashup/Clients/ Mast/Portal.html}}. We used the simple aperture photometry (SAP) light curves in the current work. We visually inspected the TESS light curves of approximately 2000 EBs present in a bibliography maintained by the first author\footnote{A hobby during Covid-19 lockdowns.}, and identified four objects that show both eclipses and previously-unrecognised $\gamma$\,Dor pulsations: CM\,Lac, MZ\,Lac, RX\,Dra and V2077\,Cyg.

CM\,Lac was observed only in sector 16 (2019/09/11 to 2019/10/07) and one further sector of observations is scheduled\footnote{\texttt{https://heasarc.gsfc.nasa.gov/cgi-bin/tess/ webtess/wtv.py}} (sector 56, 2022 September). MZ\,Lac was observed in two consecutive sectors, 16 and 17 (2019/11 to 2019/11/02) and this will recur in sectors 56--57 (2022 Sept-Oct). By contrast, extensive observations exist for RX\,Dra because it is sited within the northern continuous viewing zone (CVZ) of TESS. It was observed in sectors 14--26 (2019/07/18 to 2020/07/04) and 40--41 (2021/06/08 to 2021/08/20) and observations are currently being obtained in sectors 47--60 (2021/12/30 to 2023/01/18). V2077\,Cyg was observed in sectors 14 (2019/07/18 to 2019/08/15), 26 (2020/06/08 to 2020/07/04) and 41 (2021/07/23 to 2021/08/20), and an additional four consecutive sectors (53--56) will be observed in 2022. The light curves for one sector of each of the four targets are shown in Fig.\,\ref{fig:time}.

Following the target selection, we carefully reassessed the quality of the reduced SAP light curves. First, we inspected the aperture masks used in the pixel data and evaluated both the level of captured flux and the level of contamination from nearby stars. For all stars, the standard SPOC aperture mask was accepted, with the exception of RX\,Dra. Here, we re-extracted the light curve from the pixel data using custom aperture masks, to account for the varying contamination level between sectors by the nearby non-variable star 2MASS~J19024829+5844050. Second, we applied additional detrending to the light curves by fitting low-order polynomials. To avoid overfitting, these detrending curves were optimised simultaneously with preliminary binary and pulsation models, built from a sum of sine waves:
\begin{eqnarray*}
f(t) & = & \sum_{i=1}^{10} a_i\sin\left(2\pi\nu_i \left(t - t_0\right) + \phi_i\right) \\
     &   & + \sum_{j=1}^{10}a_j\sin\left(2\pi j\nu_{\rm orb} \left(t - t_0\right) + \phi_j\right)
\end{eqnarray*}
Here, $\nu_i$ and $\nu_{\rm orb}$ are estimates for the dominant pulsation frequencies and the orbital frequency, $a_i$ and $a_j$ are their amplitudes, and $\phi_i$ and $\phi_j$ are their phases, respectively. The mean timestamp of each light curve was used as the zeropoint $t_0$.


\begin{table*} \centering
\caption{\label{tab:jktebop} Parameters measured from the eclipses using the {\sc jktebop}
code. Parameters with a superscripted N were calculated using the nominal physical constants
and solar quantities defined by the IAU \citep{Prsa+16aj}. \reff{The \Teff\ values are from
\citet{LiakosNiarchos12apss} for CM\,Lac and from \citet{Molenda+07aca} for V2077\,Cyg.}}
\setlength{\tabcolsep}{4pt}
\begin{tabular}{lcccc}
\hline
\hline
                            & CM Lac                        & MZ Lac                      & RX Dra                        & V2077 Cyg                   \\
\hline
\multicolumn{5}{l}{Fitted parameters:} \\
$r_{\rm A}+r_{\rm B}$       & $0.35383 \pm 0.00078$         & $0.2557 \pm 0.0024$         & $0.21854 \pm 0.000099$        & $0.12303 \pm 0.00022$       \\
$k$                         & $0.902 \pm 0.018$             & $0.893 \pm 0.027$           & $0.7144 \pm 0.0010$           & $0.4933 \pm 0.0063$         \\
$i$ ($^\circ$)              & $87.573 \pm 0.082$            & $88.53 \pm 0.41$            & $88.589 \pm 0.018$            & $88.668 \pm 0.048$          \\
$J$                         & $0.5956 \pm 0.0074$           & $0.888 \pm 0.048$           & $0.8063 \pm 0.0055$           & $0.321 \pm 0.013$           \\
$L_3$                       & $-0.0043 \pm 0.0070$          & $0.377 \pm 0.018$           & $0.0176 \pm 0.0013$           & $-0.026 \pm 0.018$          \\
$u_{\rm A}$                 & $0.237 \pm 0.029$             & $0.62 \pm 0.13$             & $0.300 \pm 0.031$             & $0.248 \pm 0.033$           \\
$u_{\rm B}$                 & $0.182 \pm 0.040$             & $0.673 \pm 0.073$           & $0.195 \pm 0.057$             & $0.35 \pm 0.11$             \\
$v_{\rm A}$                 & 0.232 (fixed)                 & 0.229 (fixed)               & $0.081 \pm 0.050$             & 0.229 (fixed)               \\
$v_{\rm B}$                 & 0.229 (fixed)                 & 0.229 (fixed)               & $0.316 \pm 0.096$             & 0.229 (fixed)               \\
$e\cos\omega$               & $0.000289 \pm 0.000082$       & $0.00732 \pm 0.00025$       & $0.000030 \pm 0.000010$       & 0.0 fixed                   \\
$e\sin\omega$               & $0.0009 \pm 0.0023$           & $0.4116 \pm 0.0079$         & $0.00094 \pm 0.00062$         & 0.0 fixed                   \\
$P$ (d)                     & $1.6046879 \pm 0.0000046$     & $3.158794 \pm 0.00009$      & $3.78639491 \pm 0.00000009$   & $5.937226 \pm 0.000068$     \\
$T_0$ (BJD/TDB)             & $2458752.67073 \pm 0.00032$   & $2458765.08683 \pm 0.00014$ & $2458857.410151 \pm 0.000002$ & $2458697.52198 \pm 0.00028$ \\
$K_{\rm A}$ (\kms)          & $120.0 \pm 3.4$               &                             &                               & $54.52 \pm 0.82$            \\
$K_{\rm B}$ (\kms)          & $157.0 \pm 3.3$               &                             &                               &                             \\
$V_{\rm\gamma,A}$ (\kms)    & $22.7 \pm 1.9$                &                             &                               & $-0.23 \pm 0.40$            \\
$V_{\rm\gamma,B}$ (\kms)    & $25.0 \pm 1.8$                &                             &                               &                             \\
\multicolumn{5}{l}{Derived parameters:} \\
$r_{\rm A}$                 & $0.1860 \pm 0.0015$           & $0.1351 \pm 0.0032$         & $0.12747 \pm 0.00008$         & $0.08238 \pm 0.00038$       \\
$r_{\rm B}$                 & $0.1678 \pm 0.0021$           & $0.1296 \pm 0.0016$         & $0.09107 \pm 0.00008$         & $0.04064 \pm 0.00034$       \\
$e$                         & $0.0009 \pm 0.0015$           & $0.4117 \pm 0.0079$         & $0.00094 \pm 0.00059$         &                             \\
$\omega$ ($^\circ$)         & $71^{+213}_{-14}$             & $88.091 \pm 0.041$          & $88.2 \pm 1.5$                &                             \\
Light ratio                 & $0.495 \pm 0.018$             & $0.691 \pm 0.062$           & $0.4088 \pm 0.0008$           & $0.0750 \pm 0.0022$         \\
Mass of star A (\Msunnom)   & $2.01 \pm 0.10$               &                             &                               &                             \\
Mass of star B (\Msunnom)   & $1.54 \pm 0.09$               &                             &                               &                             \\
Radius of star A (\Rsunnom) & $1.636 \pm 0.032$             &                             &                               &                             \\
Radius of star B (\Rsunnom) & $1.476 \pm 0.031$             &                             &                               &                             \\
Surface gravity of star A   & $4.314 \pm 0.011$             &                             &                               &                             \\
Surface gravity of star B   & $4.286 \pm 0.016$             &                             &                               & $4.607 \pm 0.007$           \\
\Teff\ of star A (K)        & $8700 \pm 300$                &                             &                               & $7066 \pm 201$              \\
\Teff\ of star B (K)        & $7034 \pm 270$                &                             &                               &                             \\
\hline
\end{tabular}
\end{table*}

\section{Analysis methods}
\label{sec:anal}

\subsection{Light curve analysis}
\label{sec:anal:lc}

All four systems are well-detached EBs so are suitable for analysis with the {\sc jktebop} code\footnote{\texttt{http://www.astro.keele.ac.uk/jkt/codes/jktebop.html}} \citep{Me++04mn2,Me13aa}. In this code, the fractional radii of the stars\footnote{$r_{\rm A} = \frac{R_{\rm A}}{a}$ and $r_{\rm B} = \frac{R_{\rm B}}{a}$ where $R_{\rm A}$ and $R_{\rm B}$ are the true radii and $a$ is the semimajor axis of the relative orbit.} are parameterised as their sum ($r_{\rm A}+r_{\rm B}$) and ratio ($k = \frac{r_{\rm B}}{r_{\rm A}}$), and the orbital eccentricity ($e$) and argument of periastron are parameterised using the combination terms $e\cos\omega$ and $e\sin\omega$. Other fitted parameters include the orbital period ($P$), midpoint of primary eclipse ($T_0$), orbital inclination ($i$), and the central surface brightness ratio ($J$). We used the quadratic limb darkening (LD) law, fitted for the linear LD coefficients ($u_{\rm A}$ and $u_{\rm B}$), and fixed the quadratic LD coefficients ($v_{\rm A}$ and $v_{\rm B}$) to theoretical values \citep{Claret18aa}. We define the primary eclipse to be the deeper of the two types of eclipse, the primary star (the star eclipsed during primary eclipse) to be star A, and the secondary to be star B.

The presence of pulsations complicates the light curve analysis. In our initial analysis we ignored this and fitted the entire light curve with a {\sc jktebop} model including one low-order polynomial versus time per TESS half-sector to remove slow drifts in brightness due to instrumental or astrophysical effects. Once the residuals of the fit were obtained, these were subsequently used in the pulsation analysis (see below). For CM\,Lac and V2077\,Cyg we then extracted each eclipse from the data, plus half an eclipse duration either side, and then fitted these together with a low-order polynomial for each eclipse representing the pulsation-induced brightness changes around the time of the eclipse. This procedure was successful in the case of CM\,Lac, but led to small systematic biases for V2077\,Cyg due to the amplitude and complexity of the pulsational signature. For RX\,Dra the surfeit of data mandated a different approach: an initial fit to the full data followed by phase-binning down into a more manageable number of observations (see Section\,\ref{sec:rx}).

Uncertainties in the fitted parameters were calculated using the Monte Carlo and residual-permutation simulations implemented in {\sc jktebop} \citep{Me08mn}. The Monte Carlo simulations account for correlations between photometric parameters, and required the errorbars in the TESS data to be rescaled to force a reduced $\chi^2$ of $\chi^2_\nu = 1$. The residual-permutation simulations are sensitive to red noise in data, so are useful in the case of unmodelled pulsations. The two errorbars were checked for every parameter and the larger of the two retained in each case. From recent analyses of the TESS light curves of other EBs \citep{Me21obs3,Me21obs4} we have found that the Monte Carlo and residual-permutation simulations typically agree with each other and with errorbars obtained from splitting up light curves into multiple subsets for analysis in isolation. Additional support for the reliability of the Monte Carlo and residual-permutation errorbars, and of the {\sc jktebop} model, comes from a recent analysis of AI\,Phe by multiple researchers working independently and using multiple codes and error estimation methods \citep{Maxted+20mn}. The systematics in the modelling of V2077\,Cyg have only a small effect on our results because the system is totally-eclipsing and thus possesses a higher intrinsic determinacy of the photometric parameters \citep{Kopal59book}.

\subsection{Pulsation analysis}
\label{sec:anal:puls}

After a good fit to the light curve had been obtained, the residuals of the fit were calculated and subjected to a frequency analysis. The data points that were obtained during the eclipses were excluded to minimise the influence of any remaining binary signal in the residuals. Because the available TESS data are in most cases limited to one or two sectors per star, which are generally insufficient to resolve individual g-mode pulsations, this remains a preliminary analysis. Using a Lomb-Scargle periodogram \citep{Scargle82apj}, we iteratively prewhitened the light curve \citep{Degroote+09aa} to measure the pulsations for which the signal-to-noise ratio S/N $\geqslant$ 4 \citep{Breger+93aa}.

The amplitudes, frequencies and phases of these pulsations were subsequently optimised with a non-linear least-squares fit \citep{Bowman17book,BowmanMichielsen21aa} and the S/N ratios were re-evaluated using these optimised parameter values. The derived pulsation frequencies and their combinations were then compared with harmonics of the binary orbital frequency to identify possible tidally excited \citep[e.g.,][]{Fuller17mn,Guo+19apj}, tidally perturbed \citep[e.g.,][]{ReyniersSmeyers03aa,Bowman+19apj,Steindl++21aa} or non-linearly coupled oscillations \citep[e.g.,][]{Burkart+12mn,Guo20apj}.

Finally, we searched for combination frequencies
$$ \nu_k = n_i\nu_i + n_j\nu_j $$
where $\nu_i$, $\nu_j$ and $\nu_k$ are measured frequencies, and $n_i$ and $n_j$ are integer numbers such that $|n_i| + |n_j|\leq 2$. Such combination frequency identifications were accepted if they agreed within 3$\sigma$.

\subsection{Physical properties}
\label{sec:anal:phys}

Published spectroscopic orbits exist for both components of the CM\,Lac system, allowing us to measure its physical properties directly. To do so we included the published radial velocities (RVs) in the {\sc jktebop} analysis in order to determine the absolute masses and radii of the stars. In the case of V2077\,Cyg RVs are available for the primary but not the secondary component. In this situation the masses and radii of the stars cannot be measured directly \citep[e.g.][]{Hilditch01book}, but it is possible to obtain the surface gravity of the secondary \citep{Me+04mn3,Me++07mn} and thus verify its evolutionary status. The measured properties of the four targets are given in Table\,\ref{tab:jktebop}, including published effective temperature (\Teff) values where available.


\begin{figure*}
\includegraphics[width=\textwidth,angle=0]{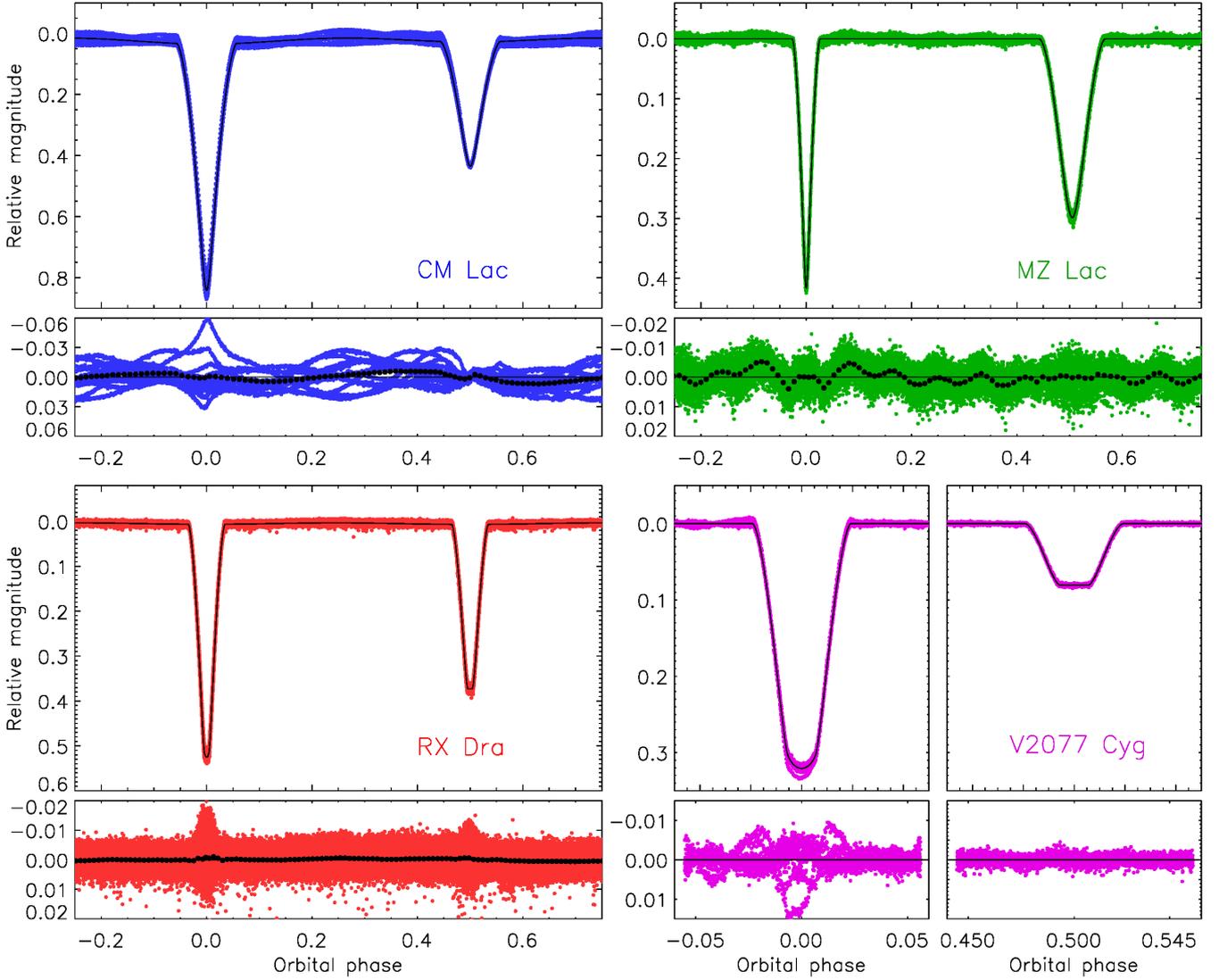}
\caption{\label{fig:phase} \reff{TESS SAP light curves of the four EBs plotted versus orbital phase.
In each case the coloured points give the observations and the black line shows the best fit. The
residuals are plotted below each main panel on an inflated scale. Black points show the residuals
binned according to orbital phase. The plot for V2077\,Cyg has been split into panels concentrating
on the eclipses for clarity. The polynomials versus time have been subtracted from the observations
in all cases to remove any slow trends in brightness that would otherwise blur the figure in the
vertical direction.}} \end{figure*}

\section{Discussion of individual systems}
\label{sec:each}

\subsection{CM Lacertae}
\label{sec:cm}


\subsubsection{Analysis of the binarity}

CM\,Lac was discovered to be an eclipsing binary by \citet{Wachmann31an} and physical properties were measured by \citet{Popper68apj}. The most detailed study of the system was by \citet{LiakosNiarchos12apss}, who determined its physical properties based on light curves observed in the $BVRI$ bands and RVs from a set of 28 medium-resolution spectra.

The TESS light curve of CM\,Lac (Fig.\,\ref{fig:time}) contains 12 primary and 14 secondary eclipses. The pulsations are strong enough to significantly affect the eclipse depths so we cut out the data around each eclipse from the light curve and modelled the out-of-eclipse brightness through each as a quadratic function, as described in Section\,\ref{sec:anal:lc}. We found that it was possible to constrain one LD coefficient for each star. We also fitted for the amount of third light ($L_3$), obtaining a slightly negative value. This is physically plausible if the background subtraction in the TESS images is too strong. Although the orbit appears circular, a slight eccentricity is preferred for the TESS data. A phased light curve is shown in Fig.\,\ref{fig:phase}.

\begin{figure} \includegraphics[width=\columnwidth,angle=0]{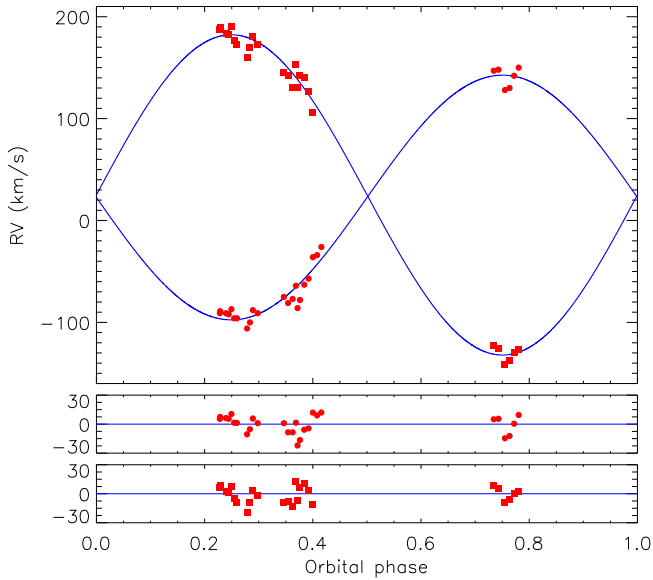}
\caption{\label{fig:cm:rv} \reff{Fitted spectroscopic orbit for CM\,Lac (blue lines)
compared to the measured RVs from \citet{LiakosNiarchos12apss}: red circles for star~A
and red squares for star~B. The residuals of the fit are shown in the lower panels.}} \end{figure}

Our final solution includes the RVs from \citet{LiakosNiarchos12apss} in order to determine the full properties of the system; we fitted for the velocity amplitudes ($K_{\rm A}$ and $K_{\rm B}$) and systemic velocities ($V_{\rm\gamma,A}$ and $V_{\rm\gamma,B}$) of both stars \reff{(Fig.\,\ref{fig:cm:rv})}. The reduced $\chi^2$ of the fit was forced to unity for each dataset by scaling the data errors. We tried including historical times of primary eclipse as well \citep{Wachmann31an,Kreiner++01book} but they occurred systematically too late so we rejected them -- this may suggest that the orbital period of CM\,Lac is not constant. The uncertainties from residual permutation are larger than those from Monte Carlo. Our results are in Table\,\ref{tab:jktebop}: they agree well with those of \citet{LiakosNiarchos12apss}. Further RVs are needed to refine the mass measurements, which are currently precise to only 5\%.

\subsubsection{Analysis of the pulsations}

\begin{figure} \includegraphics[width=\columnwidth]{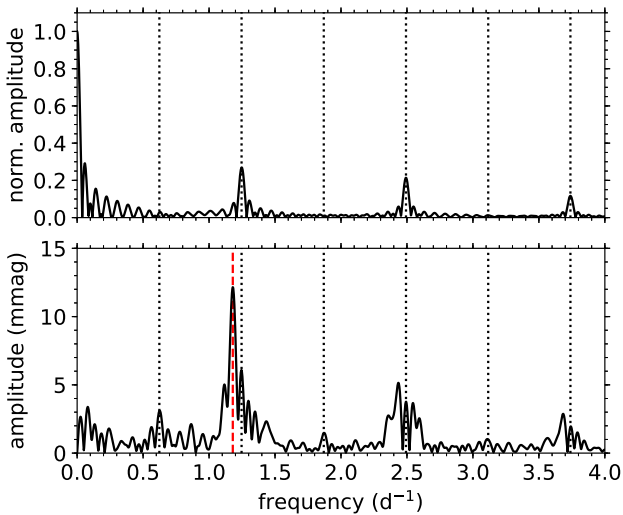}
\caption{\label{fig:cmlac_lombscargle} \reff{Spectral window (top panel) and} part of
the Lomb-Scargle periodogram (black) of the TESS light curve of CM\,Lac, excluding the
eclipses. The harmonics of the orbital frequency $\nu_{\rm orb}$ (dotted grey lines) and
the extracted pulsation frequency (dashed red line) are also indicated.} \end{figure}

Our frequency analysis reveals the presence of $\gamma$\,Dor pulsations. This can easily be seen from the Lomb-Scargle periodogram of the TESS light curve (excluding the eclipses) shown in Fig.\,\ref{fig:cmlac_lombscargle}. There is unresolved variability with frequencies between 1.0 and 1.5\,$\rm d^{-1}$. This is most likely caused by g-mode pulsations, as the (tentative) dominant frequency that can be obtained with iterative prewhitening (listed in Table\,\ref{tab:cmlac_freq}), differs significantly from the harmonics of the binary orbital frequency. However, because the variability is unresolved, the measured frequency value cannot be assigned to a single g-mode pulsation. The asymmetric shape of the light curve is also typical for g-mode pulsations \citep{Kurtz+15mn}.

\reff{We note that the second harmonic of the binary orbital frequency also lies within the frequency range of the observed variability, as shown in Fig.\,\ref{fig:cmlac_lombscargle}. However, because the available TESS data only cover 27\,d, we cannot investigate the nature of the variability at this frequency. It may be caused by a residual signature of the binarity or unresolved nearby g-mode pulsation frequencies, but more high-precision photometric data are needed to resolve this issue. The unresolved variability around the fourth and sixth harmonics of the orbital frequency are likely aliases of the unresolved variability around the second harmonic, as can be deduced from the spectral window shown in the top panel of Fig.\,\ref{fig:cmlac_lombscargle}. From a visual inspection of the TESS light curve, we find that the pulsations remain visible during the primary eclipse, but vanish during the secondary eclipse. This means that the g-mode pulsations belong to the secondary star.}

\begin{table} \centering
\caption{\label{tab:cmlac_freq} The parameter values for the significant prewhitened frequency of
CM\,Lac. Bracketed quantities indicate the uncertainty in the final digit of the preceding number.}
\begin{tabular}{cccc}
\hline \hline
Frequency ($\rm d^{-1}$) & Amplitude (mmag) & Phase (rad) & S/N \\
\hline
1.1799 (2) & 12.5 (1) &  0.4276 (12) & 6.0 \\
\hline
\end{tabular}
\end{table}


\subsection{MZ Lacertae}
\label{sec:mz}


\begin{figure} \includegraphics[width=\columnwidth]{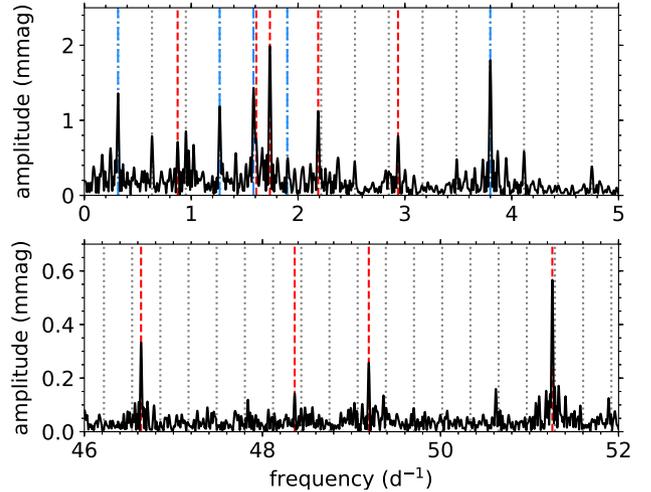}
\caption{\label{fig:mzlac_lombscargle} Sections of the Lomb-Scargle periodogram (black) of the
residual light curve of MZ\,Lac.
The harmonics of the orbital frequency $\nu_{\rm orb}$ (dotted grey lines) and the extracted
pulsation frequencies are also indicated. The dashed red lines and dash-dotted blue lines mark
non-harmonic and harmonic pulsation frequencies, respectively.} \end{figure}

\begin{table*}
\caption{\label{tab:mzlac_freq} An overview of the parameter values for the significant frequencies of MZ\,Lac. For
frequencies that lie close to harmonics of the binary orbital frequency $\nu_{\rm orb}$, the differences between those
frequencies and their corresponding harmonics are also given. Possible combination frequencies  are listed in the last column.}
\begin{tabular}{llcrrrrl}
\hline \hline
 & Frequency ($\rm d^{-1}$) & Amplitude (mmag) & Phase (rad) & S/N & $i$ & $i\nu_{\rm orb} - \nu$ & Comments \\
\hline 
$\nu_1$    &  0.31523 (14)  &  1.66 (3)  &  $ 0.495$  (2)  &  12.3  &  1 & $-$0.00135 (14) &                                                \\
$\nu_2$    &  0.8729 (4)    &  0.70 (2)  &  $-0.477$  (5)  &   5.9  &    &                 &                                                \\
$\nu_3$    &  1.2665 (2)    &  1.46 (3)  &  $ 0.131$  (3)  &  14.7  &  4 &    0.0002   (2) &                                                \\
$\nu_4$    &  1.5819 (4)    &  0.60 (3)  &  $-0.103$  (7)  &   6.3  &  5 & $-$0.0010   (4) &                                                \\
$\nu_5$    &  1.6094 (3)    &  0.74 (2)  &  $-0.326$  (5)  &   7.9  &    &                 &                                                \\
$\nu_6$    &  1.73674 (14)  &  1.86 (2)  &  $ 0.424$  (2)  &  20.5  &    &                 & $2\nu_2 + \nu_6 = 11.000(3)\,\nu_{\rm orb}$    \\
$\nu_7$    &  1.8991 (4)    &  0.74 (3)  &  $-0.441$  (5)  &   8.2  &  6 & $-$0.0004   (4) &                                                \\
$\nu_8$    &  2.1884 (2)    &  1.13 (2)  &  $ 0.038$  (3)  &  14.2  &    &                 & $\nu_{5} + \nu_8 = 11.9966(12)\,\nu_{\rm orb}$ \\
$\nu_9$    &  2.9357 (3)    &  0.85 (2)  &  $-0.458$  (4)  &  12.8  &    &                 & $\nu_2 + \nu_9 = 12.0303(16)\,\nu_{\rm orb}$   \\
$\nu_{10}$ &  3.79908 (14)  &  1.81 (2)  &  $ 0.424$  (2)  &  32.5  & 12 &    0.0002   (2) &                                                \\
$\nu_{11}$ &  46.6367 (7)   &  0.33 (2)  &  $ 0.264$ (11)  &  13.0  &    &                 &                                                \\
$\nu_{12}$ &  48.362 (2)    &  0.14 (2)  &  $-0.20$   (3)  &   4.4  &    &                 &                                                \\
$\nu_{13}$ &  49.1942 (9)   &  0.26 (2)  &  $-0.408$ (14)  &   8.4  &    &                 &                                                \\
$\nu_{14}$ &  51.2559 (4)   &  0.57 (2)  &  $-0.334$  (7)  &  20.3  &    &                 &                                                \\
\hline
\end{tabular}
\end{table*}

\subsubsection{Analysis of the binarity}

MZ\,Lac was found to be an EB by \citet{MillerWachmann71ra}, where it is labelled VV\,399, but has been the subject of very little study since. It exhibits apsidal motion \citep{Silhan90jaavso} with a period of $424 \pm 6$\,yr \citep{Bulut+16aipc}. The TESS light curve shows deep eclipses (Fig.\,\ref{fig:time}) and covers two sectors. The secondary eclipse is at approximately phase 0.5 but is significantly longer than the primary, so $e\cos\omega \approx 0$ but $e\sin\omega$ is large.

The pulsations in MZ\,Lac have a much lower amplitude than the eclipses so we treated them as red noise: we fitted the whole dataset with {\sc jktebop} without attempting to reject out-of-eclipse data or compensate for the pulsations in any way. We fitted for the same parameters as for CM\,Lac, except for those related to RVs and for the inclusion of only four polynomials for the out-of-eclipse brightness of the system \reff{(one for each TESS half-sector)}. Due to the apsidal motion we did not attempt to use published eclipse timings to help constrain the orbital ephemeris. A substantial amount of third light is required to obtain a good fit to the data.

We were able to find a good fit to the light curve (Table\,\ref{tab:jktebop}). The residual-permutation errorbars are significantly larger than the Monte Carlo errorbars, which suggests it would be possible to improve the light curve analysis in future by either subtracting the pulsations or fitting for them simultaneously. The full physical properties of the system cannot be calculated as no RVs are available. \textit{Gaia} DR2 \citep{Gaia18aa} gives a \Teff\ of the system of 6900\,K; if this is appropriate for star A then the \Teff\ of star B is 6700\,K based on the value of $J$ we measure. These values are consistent with the presence of $\delta$\,Sct and $\gamma$\,Dor pulsations \citep{Grigahcene+10apj,BowmanKurtz18mn,Murphy+19mn}.


\subsubsection{Analysis of the pulsations}

From a detailed analysis of the TESS photometry, we find that MZ\,Lac exhibits an interesting variety of pulsation behaviour. Following the binary modelling, we excluded the eclipses from the residual light curve and used this clipped light curve for the iterative prewhitening. The prewhitened pulsation frequencies of MZ\,Lac are listed in Table\,\ref{tab:mzlac_freq} and shown in Fig.\,\ref{fig:mzlac_lombscargle}.

Ten frequencies below 5\,$\rm d^{-1}$ have been measured from the TESS light curve. Five of these lie significantly close to harmonics of the orbital frequency\footnote{\refff{MZ\,Lac experiences apsidal motion, so its sidereal and anomalistic periods are not identical. However, they differ by only 0.000035\,d so our analysis is not affected by which we choose to represent the orbital frequency.}}, and four are well within 3$\sigma$. A visual inspection of the light curve, phase-folded with the binary orbital period as shown in Fig.\,\ref{fig:phase}, reveals that these frequencies are likely related to tidally excited pulsations. An imperfect fit of the binary orbit to the light curve might cause similar residual signal in the Fourier transform at integer multiples of the orbital frequency. However, the dominant variation seen in the phase-folded light curve (in Fig.\,\ref{fig:phase}) occurs on a timescale much shorter than the orbital period, which is consistent with tidally excited pulsations, but not with residual signal of the binary orbit itself.

We also obtain five non-harmonic low frequencies. While these have values of typical $\gamma$\,Dor-type pulsations, three pairs of them form combination frequencies that (almost) coincide with harmonics of the orbital frequency:
$2\nu_{2} + \nu_{6} = 11.000(3)\,\nu_{\rm orb},$ $\nu_{5} + \nu_8  = 11.9966(12)\,\nu_{\rm orb}$ and $\nu_2 + \nu_9 = 12.0303(16)\,\nu_{\rm orb}$. The first two combinations match the closest orbital harmonics within $3\sigma$, and the last one matches the orbital harmonic within $\nu_{\rm res}$. This can be caused by nonlinear mode coupling, whereby the orbital harmonic is the parent frequency and the observed non-harmonic frequencies are the child frequencies \citep[e.g.,][]{Burkart+12mn,Weinberg++13apj,Guo++17apj}.  Finally, there are four pressure-mode (p-mode) pulsations, with frequencies between 46 and 51.5\,$\rm d^{-1}$.

From a visual inspection of the TESS light curve shown in Fig.\ref{fig:phase}, we find that the pulsations appear to be less visible during \emph{both} eclipses. Our best explanation for this apparent physical inconsistency is that the (tidally excited) pulsations were partially fitted by the binary model in the eclipses. This could have happened in particular if the pulsations had a similar effect on the eclipse shape as LD, as some LD coefficients were included as fitted parameters. The fitted LD coefficients are indeed larger than expected. Whilst one approach to fixing this problem might be to fix the LD coefficients, this risks biasing the measured radii because theoretical LD coefficients are not completely reliable. An alternative approach in future could be to simultaneously model the eclipses and the pulsations.


\subsection{RX Draconis}
\label{sec:rx}

\begin{figure*} \includegraphics[width=\textwidth]{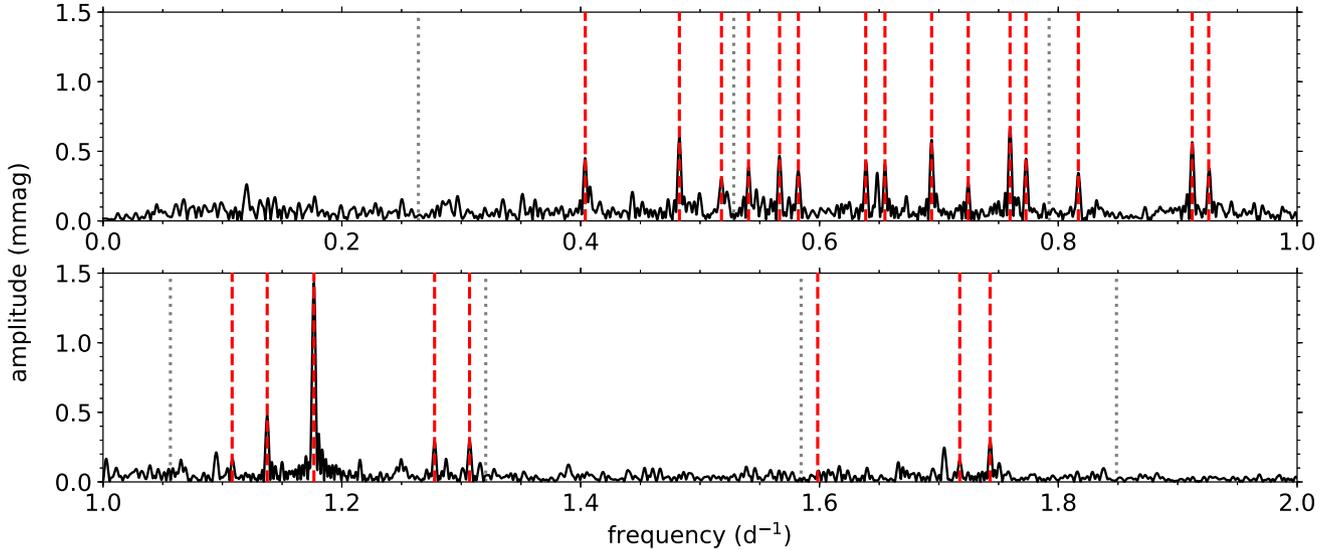}
\caption{\label{fig:rxdra_lombscargle} Sections of the Lomb-Scargle periodogram
(black) of the residual light curve of RX\,Dra. The harmonics of the orbital frequency
$\nu_{\rm orb}$ (dotted grey lines) and the extracted pulsation frequencies
(dashed red lines) are also indicated.} \end{figure*}


\begin{table} \centering
\caption{\label{tab:rxdra_freq} An overview of the parameter values for the significant
frequencies of RX\,Dra. Possible combination frequencies within 3$\sigma$ are also listed.}
\setlength{\tabcolsep}{4pt}
\begin{tabular}{llcrrl}
\hline\hline
           & Frequency      & Amplitude   &     Phase       &  S/N   & Combinations          \\
           & ($\rm d^{-1}$) & (mmag)      &     (rad)       &        &                       \\
\hline
$\nu_{1}$  &  0.40386  (2)  &  0.407 (5)  &  $0.104 $  (2)  &   9.4  &                       \\
$\nu_{2}$  &  0.48268  (1)  &  0.64  (5)  &  $-0.486$  (1)  &  15.2  &                       \\
$\nu_{3}$  &  0.51792  (3)  &  0.252 (5)  &  $-0.484$  (3)  &   6.0  & $\nu_{19} - \nu_{11}$ \\
$\nu_{4}$  &  0.54059  (2)  &  0.376 (5)  &  $0.296 $  (2)  &   9.1  &                       \\
$\nu_{5}$  &  0.56659  (2)  &  0.516 (5)  &  $0.492 $  (2)  &  12.5  &                       \\
$\nu_{6}$  &  0.58234  (2)  &  0.395 (5)  &  $0.371 $  (2)  &   9.6  &                       \\
$\nu_{7}$  &  0.63871  (2)  &  0.41  (5)  &  $0.235 $  (2)  &  10.1  &                       \\
$\nu_{8}$  &  0.65485  (2)  &  0.373 (5)  &  $0.491 $  (2)  &   9.2  & $\nu_{17} - \nu_{2}$  \\
$\nu_{9}$  &  0.69392  (1)  &  0.596 (5)  &  $0.204 $  (1)  &  14.8  & $\nu_{18} - \nu_{2}$  \\
$\nu_{10}$ &  0.72459  (4)  &  0.224 (5)  &  $0.401 $  (4)  &   5.6  & $\nu_{20} - \nu_{6}$  \\
$\nu_{11}$ &  0.75959  (1)  &  0.674 (5)  &  $-0.414$  (1)  &  17.0  &                       \\
$\nu_{12}$ &  0.77293  (2)  &  0.397 (5)  &  $-0.096$  (2)  &  10.0  &                       \\
$\nu_{13}$ &  0.81678  (2)  &  0.365 (5)  &  $0.391 $  (2)  &   9.3  &                       \\
$\nu_{14}$ &  0.91214  (2)  &  0.526 (5)  &  $-0.26 $  (2)  &  13.8  &                       \\
$\nu_{15}$ &  0.92597  (2)  &  0.393 (5)  &  $0.115 $  (2)  &  10.4  &                       \\
$\nu_{16}$ &  1.10811  (5)  &  0.179 (5)  &  $-0.442$  (5)  &   5.0  &                       \\
$\nu_{17}$ &  1.13749  (2)  &  0.466 (5)  &  $0.479 $  (2)  &  13.2  &                       \\
$\nu_{18}$ &  1.176625 (6)  &  1.405 (5)  &  $-0.230$3 (6)  &  41.0  &                       \\
$\nu_{19}$ &  1.27764  (3)  &  0.276 (5)  &  $0.265 $  (3)  &   8.5  &                       \\
$\nu_{20}$ &  1.30693  (3)  &  0.274 (5)  &  $-0.002$  (3)  &   8.6  &                       \\
$\nu_{21}$ &  1.5985   (6)  &  0.132 (5)  &  $-0.03 $  (6)  &   5.1  &                       \\
$\nu_{22}$ &  1.71749  (4)  &  0.187 (5)  &  $0.324 $  (4)  &   7.7  &                       \\
$\nu_{23}$ &  1.74289  (3)  &  0.289 (5)  &  $-0.44 $  (3)  &  12.1  &                       \\
$\nu_{24}$ &  2.8467   (1)  &  0.054 (5)  &  $0.37  $  (1)  &   4.3  &                       \\
\hline
\end{tabular}
\end{table}

\subsubsection{Analysis of the binarity}

RX\,Dra has been known to be eclipsing for over a century. The first photometric analysis was given by \citet{Shapley13apj} but it has received very little attention subsequently. It came to the authors' attention by its inclusion in a list of EBs that are candidates for hosting pulsating stars \citep{Soydugan+06mn}, and inspection of its TESS light curve revealed clear pulsations of the $\gamma$\,Dor type. TESS observations are available for 13 consecutive sectors (14--26) followed by two further sectors (40 and 41). At the time of writing, observations are also ongoing for 14 consecutive sectors (47--60); see Section\,\ref{sec:obs}. We base our results in the current work on the data from sectors 14--26.

Because of the huge amount of data (230\,766 observations sampled at 120\,s cadence for one year) covering a large number of eclipses (approximately 170) we first obtained a preliminary solution with {\sc jktebop} to establish the orbital ephemeris. We then converted the data into orbital phase and binned them into 1000 points equally spaced in phase. The primary eclipse is annular and the secondary eclipse is total. A good fit to the binned data can be obtained comparatively easily due to the total eclipses. We included orbital eccentricity and third light as fitted parameters and both come out at very small but probably non-zero. It was also possible to fit for \emph{both} LD coefficients for both stars, an occurrence the first author has not previously encountered despite having recently studied the TESS light curves of over 50 EBs. The errorbars from the Monte Carlo algorithm were adopted as they are significantly larger than those from the residual-permutation approach. Our results are given in Table\,\ref{tab:jktebop}. The uncertainties in the fractional radii are below 0.1\%, but we recommend that interested users increase them to at least 0.1\% as this is the lower limit of reliability established by comparing independent analyses of the same data for a similar EB \citep{Maxted+20mn}.

\subsubsection{Analysis of the pulsations}

RX\,Dra is an outlier in our sample. While we only have one or two TESS sectors of data available for most of our stars, RX\,Dra is located in the northern CVZ and has been observed during 13 consecutive \reff{sectors}, spanning 351 days in total. This has allowed us to better resolve the individual stellar pulsation frequencies and measure them with a higher precision, as listed in Table~\ref{tab:rxdra_freq} and shown in Fig.\,\ref{fig:rxdra_lombscargle}.

Despite the short orbital period, all observed pulsations were found to be g-mode pulsations. Four of the 24 measured frequencies form combinations within 3$\sigma$. Additionally, the observed pulsation spectrum is relatively sparse, given that we have almost a full year of photometry at our disposal. The inclusion of additional photometric observations, such as those that are currently being taken during TESS cycle 4, may improve the window functions of the stellar pulsations sufficiently to allow us to measure additional g-mode pulsation frequencies.

Similarly to the pulsations observed for MZ\,Lac, we cannot assign the pulsations of RX\,Dra to a specific component. Here, the narrow triangular shapes of the eclipses do not suffice to evaluate the pulsation amplitudes as a function of the light contribution of the individual components, as illustrated in Fig.\,\ref{fig:phase}. Additionally, the measured g-mode frequencies are too sparse to detect clear g-mode period-spacing patterns, which would have allowed us to perform pulsation mode identification and select a group or groups of pulsations that originate from the same stellar component in the binary system. There is, however, clear structure present in the observed g-mode pulsation spectrum. The inclusion of the forthcoming TESS observations will be needed to detect period-spacing patterns for RX\,Dra \citep[e.g.][]{Vanreeth+15aa} if they are present in the data.


\subsection{V2077 Cygni}
\label{sec:v2077}

\subsubsection{Analysis of the binarity}

V2077\,Cyg was identified as an EB using photometry from the \textit{Hipparcos} satellite, and given its GCVS designation by \citet{Kazarovets+99ibvs} as a result. The only study of the object so far was performed by \citet{Molenda+07aca}, who obtained 29 spectra and measured the spectroscopic orbit of the primary star. The secondary is much fainter and was not detected in the spectra. The TESS light curve (Fig.\,\ref{fig:time}) shows distinctive variability characteristic of $\gamma$\,Dor pulsators, plus primary eclipses that are much deeper than the secondary eclipses. The light curve is strongly reminiscent of the prototypical system KIC\,11285625 \citep{Debosscher+13aa}, except that the eclipses are total. TESS has observed the star during three sectors (14, 26 and 41) and four further sectors of observations (53, 54, 55 and 56) are planned. The four consecutive sectors will be valuable for increasing the precision of the measured pulsation periods. The following analyses only includes sectors 14 and 26, as sector 41 was not available at the time they were performed.

\begin{figure} \includegraphics[width=\columnwidth,angle=0]{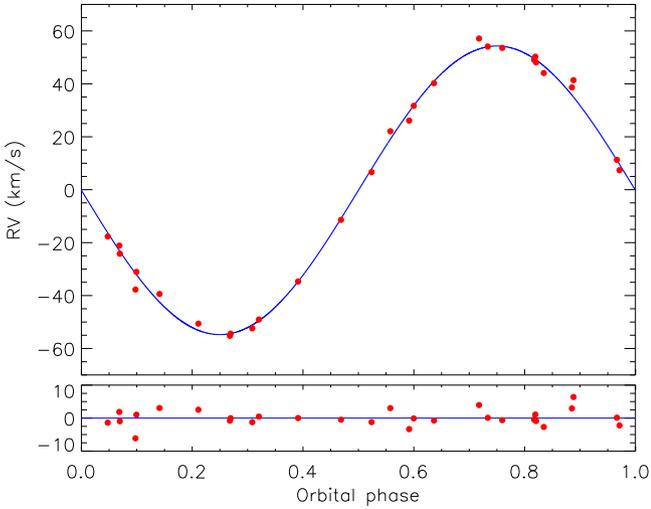}
\caption{\label{fig:v2077:rv} \reff{Fitted spectroscopic orbit for V2077\,Cyg
(blue lines) compared to the measured RVs from \citet{Molenda+07aca} for star~A
(red circles). The residuals of the fit are shown in the lower panel.}} \end{figure}

The pulsation amplitudes are comparable to the eclipse depths so we proceeded as for CM\,Lac (Sect.\,\ref{sec:cm}): each eclipse was extracted from the light curve and normalised to zero differential magnitude using a polynomial function. We found that a third-order polynomial was necessary in some cases. The light curve was then fitted as for CM\,Lac, with the exception that we had RVs for only the primary star so did not fit for $K_{\rm B}$ or $V_{\rm\gamma,B}$ \reff{Fig.\,\ref{fig:v2077:rv})}. The results are given in Table\,\ref{tab:jktebop}. The fractional radii of the stars are well-measured because of the presence of total eclipses in this system.

\citet{Molenda+07aca} measured a \Teff\ of $7066 \pm 201$\,K for the primary star, in good agreement with the \textit{Gaia} DR2 value of 6900\,K. Our value of $J$ then implies a \Teff\ of 5200\,K for the secondary star. Using equation 4 from \citet{Me++07mn} we \reff{determined} the surface gravity of the secondary star to be $\log g_{\rm B} = 4.607 \pm 0.007$, which is appropriate for a K-dwarf. The $\gamma$\,Dor pulsations arise in the primary component: the secondary is too cool to be a $\gamma$\,Dor star and it also does not contribute enough light to the system to produce pulsations with an amplitude of 0.1\,mag.

\subsubsection{Analysis of the pulsations}

\begin{figure} \includegraphics[width=\columnwidth]{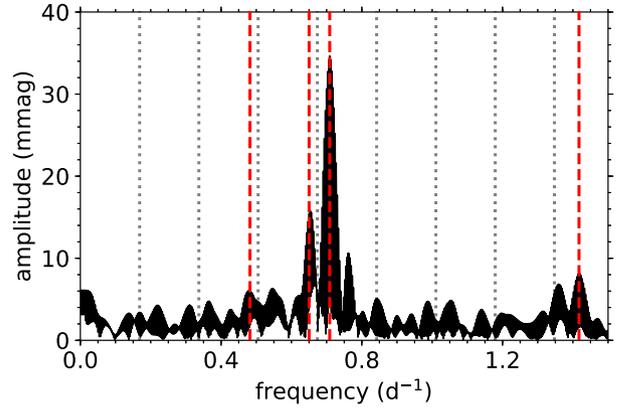}
\caption{\label{fig:v2077cyg_lombscargle} Part of the Lomb-Scargle periodogram (black)
of the residual light curve of V2077\,Cyg (after the best-fitting binary model has been
subtracted). The harmonics of the orbital frequency $\nu_{\rm orb}$ (dashed grey lines)
and the extracted pulsation frequencies (dashed red lines) are also indicated.} \end{figure}

Four significant frequencies, listed in Table\,\ref{tab:v2077cyg_freq} and illustrated in Fig.\,\ref{fig:v2077cyg_lombscargle}, can be obtained from iterative prewhitening of the residual TESS light curve. The derived parameter values for these frequencies are all consistent with those expected for $\gamma$\,Dor-type pulsations. However, the available TESS data are again insufficient to ensure that the individual g-mode pulsation frequencies are properly resolved.

More TESS data are needed to properly identify the individual pulsations in this system. Fortunately, it is scheduled to be observed in four consecutive sectors (53--56). If TESS continues to operate as expected\reff{, and precise radial velocities can be measured for both components, V2077\,Cyg could become} a useful system for the study of pulsations in a star of known mass and radius. \reff{Fig.\,\ref{fig:phase} shows an increase in residuals for the binary fit during the eclipses. This arises because the much smaller secondary star is blocking only part of the surface of the pulsating primary star and thus breaking the spherical symmetry that attenuates the observed amplitudes of pulsations. V2077\,Cyg may therefore be a good candidate for elipse mapping to spatially resolve the pulsations on the stellar surface \citep[e.g.][]{BiroNuspl11mn}.}

\begin{table} \centering
\caption{\label{tab:v2077cyg_freq} An overview of the parameter values for the significant frequencies
of V2077\,Cyg. The identification of possible combination frequencies within 3$\sigma$ is also given.}
\setlength{\tabcolsep}{5pt}
\begin{tabular}{lccccl}
\hline\hline
          & Frequency      & Amplitude &     Phase     &  S/N & Combinations           \\
          & ($\rm d^{-1}$) & (mmag)    &     (rad)     &      &                        \\
\hline
$\nu_{1}$ &  0.48209  (1)  & 6.97 (10) & $-$0.244  (2) &  4.1 &                        \\
$\nu_{2}$ &  0.65013  (1)  & 8.5   (1) & $-$0.163  (2) &  5.3 &                        \\
$\nu_{3}$ &  0.708696 (3)  & 33.4  (1) &    0.4049 (5) & 21.1 &                        \\
$\nu_{4}$ &  1.41741  (1)  & 8.72 (10) &    0.075  (2) &  6.7 & $\nu_4 \approx 2\nu_3$ \\
\hline
\end{tabular}
\end{table}


\section{Summary and conclusions}
\label{sec:conc}

The study of pulsating stars in EBs is a promising opportunity to improve our understanding of the interior physics of stars. Gravity-mode pulsations are a high priority because they can probe regions deep inside a star and thus help constrain stellar rotation, convective core boundary mixing, envelope mixing, opacity and magnetic fields (see references in Section\,\ref{sec:intro}). In pursuit of this goal we have surveyed the TESS light curves of a sample of known EBs and detected g-mode pulsations in four of these. In this work we present a preliminary analysis of all four systems.

CM\,Lac shows very deep eclipses (0.8\,mag for the primary and 0.4\,mag for the secondary) and unresolved g-modes with frequencies between $1\,\rm d^{-1}$ and $1.5\,\rm d^{-1}$ in the one sector of available TESS data. The available data are therefore insufficient for asteroseismology, but remain useful for the determination of the physical properties of stars. We measured the masses and radii of the component stars using the TESS data and the results of a published RV study, but additional RVs are needed to measure their masses to the canonical 2\% precision \citep{Andersen91aarv}.

MZ\,Lac is a more interesting system due to its large orbital eccentricity (Table\,\ref{tab:jktebop}) and richer pulsation spectrum. Based on the two sectors of TESS data for this object we found evidence for p-modes and for tidally excited or tidally perturbed g-mode pulsations. No RVs are currently available for this object so we are not yet able to determine its physical properties.

RX\,Dra has been observed for 13 consecutive sectors by TESS but this abundance of data yielded only relatively few identified frequencies. The presence of total eclipses allowed the fractional radii of the stars to be measured to exceptional precision so the system is a good candidate for precise measurement of its physical properties. RX\,Dra is in the process of being observed for a further 14 consecutive sectors by TESS and these additional data may allow the detection of further pulsation frequencies.

V2077\,Cyg shows beautiful $\gamma$\,Dor pulsations but the few currently-available TESS light curves are insufficient to perform mode identification. The system is scheduled for observation in three consecutive sectors by TESS in 2022 and these new data should greatly improve the asteroseismic potential of this system. V2077\,Cyg is not promising from the viewpoint of determining its physical properties for two reasons. The eclipses are perturbed by the high-amplitude g-mode pulsations, but are total so precise fractional radii are still measurable. The secondary star is much fainter than the primary (light ratio of 0.075 in the TESS passband) so measuring its RVs will require high-quality spectroscopy. A single-lined spectroscopic orbit exists for this system, allowing us to measure the surface gravity of the secondary component and verify that it is a normal low-mass main-sequence star.

All four systems need detailed spectroscopic study to measure their physical properties precisely, which in turn provides context for the pulsation analysis. This work has begun.


\section*{Data availability}

All data underlying this article are available in the MAST archive ({https://mast.stsci.edu/portal/Mashup/Clients/Mast/Portal.html}).

\section*{Acknowledgements}

TVR gratefully acknowledges support from the Research Foundation Flanders (FWO) under grant agreement N\textdegree 12ZB620N.
We thank Johanna Molenda-\.Zakowicz for sending us her RV measurements for V2077\,Cyg, and Dominic Bowman and the anonymous referee for discussions and comments.
The TESS data presented in this paper were obtained from the Mikulski Archive for Space Telescopes (MAST) at the Space Telescope Science Institute (STScI).
STScI is operated by the Association of Universities for Research in Astronomy, Inc., under NASA contract NAS5-26555.
Support to MAST for these data is provided by the NASA Office of Space Science via grant NAG5-7584 and by other grants and contracts.
Funding for the TESS mission is provided by the NASA Explorer Program.
This research has made use of the SIMBAD database, operated at CDS, Strasbourg, France; the SAO/NASA Astrophysics Data System; and the VizieR catalogue access tool, CDS, Strasbourg, France.


\bibliographystyle{mnras}

\bsp \label{lastpage} \end{document}